\documentclass[10pt,preprint]{aastex}

\slugcomment{To be published in the Astronomical Journal}
\received{9/18/2002}
\revised{12/19/2002}
\revised{01/07/2003}
\accepted{01/14/2003}

\def\iras{{\sl IRAS\/}}
\def\msun{M_{\odot}}

\def\lstar{L_{*}}

\def\teff{T_{\rm eff}}
\def\um{\mu {\rm m}}
\def\kp{K^{\prime}}
\makeatletter
\def\ale{\mathrel{\mathpalette\gl@align<}}
\def\age{\mathrel{\mathpalette\gl@align>}}
\def\gl@align#1#2{\lower.6ex\vbox{\baselineskip\z@skip\lineskip\z@
\ialign{$\m@th#1\hfil##\hfil$\crcr#2\crcr\sim\crcr}}}
\makeatletter

\shorttitle{Near-IR Photometry of PPN Candidates} 
\shortauthors{Ueta et al.}

\begin{document}
 
\title{Near-Infrared Photometric Survey of 
Proto-Planetary Nebula Candidates}

\author{Toshiya Ueta\altaffilmark{1,2},
Margaret Meixner\altaffilmark{1,3},
Danielle E. Moser\altaffilmark{1},
Lukasz A. Pyzowski\altaffilmark{1}, and
Jason S. Davis\altaffilmark{1,4}}

\altaffiltext{1}{Department of Astronomy, MC-221, 
University of Illinois at Urbana-Champaign, 
Urbana, IL  61801, USA;
demoser@astro.uiuc.edu,
pyzowski@astro.uiuc.edu}

\altaffiltext{2}{Current Address:
Royal Observatory of Belgium, 
Avenue Circulaire, 3,
B-1180, Brussels, Belgium;
ueta@oma.be}

\altaffiltext{3}{Current Address:
Space Telescope Science Institute, 
3700 San Martin Drive, Baltimore, MD 21218, USA;
meixner@stsci.edu}

\altaffiltext{4}{Current Address:
The Cleveland Museum of Natural History,
1 Wade Oval Drive,
University Circle,
Cleveland, OH 44106, USA;
jdavis@cmnh.org}

\begin{abstract}
We present $JHK^{\prime}$ photometric measurements 
of 78 objects mostly consisting of proto-planetary
nebula candidates.
Photometric magnitudes are determined by means of
imaging and aperture photometry.
Unlike the observations with a photometer with a
fixed-sized beam, the method of imaging photometry
permits accurate derivation of photometric values
because the target sources can be correctly identified
and confusion with neighboring sources can be
easily avoided.
Of the 78 sources observed, we report 10 cases in 
which the source seems to have been misidentified
or confused by nearby bright sources.
We also present nearly two dozen cases in which the 
source seems to have indicated a variability
which prompts a follow-up monitoring.
There are also a few sources that show previously 
unreported extendedness.
In addition, we present $H$ band finding charts of 
the target sources.
\end{abstract}


\section{Introduction\label{intro}}

Proto-planetary nebulae (PPNs) are evolved stars of 
low-to-intermediate initial mass ($0.8 - 8 \msun$) 
that are in transition from the asymptotic giant 
branch (AGB) phase to the planetary nebula (PN) phase 
of evolution (e.g., \citealt{kwok93,iben95}).
These evolved stars have an extended shell of dust 
and gas surrounding the star, which is formed by mass 
loss during the AGB phase.
Since the AGB mass loss has already ceased (or 
significantly been weakened) by the beginning of the 
PPN phase, the circumstellar shell is physically 
detached from the central star in PPNs.
This detached shell of dust acts as a reservoir
of thermal infrared (IR) emission, and therefore, 
PPNs are typically bright in the mid-IR wavelengths 
and observed as cool {\iras} sources.

Meanwhile, the PPN central stars are slightly more 
evolved than AGB stars, and their spectral types are 
usually of G to late B.
In addition, the detached circumstellar shells of 
PPN are not as optically thick as the AGB envelopes
due to expansion.
Therefore, unlike AGB stars, PPNs are often also 
bright in the optical.
This results in the ``double-peaked'' spectral energy 
distribution (SED) that is typical of PPNs \citep{veen89}.
This characteristic signature of PPNs has been the
primary means to identify these highly evolved stars
in which optical counterparts are searched at the 
positions of unidentified cool {\iras} sources
\citep{kwok93}.
Multi-color photometry has also been used to 
identify PPNs (e.g., \citealt{hrivnak89,fujii02}).

Although PPNs may be bright at the mid-IR and optical,
they are not necessarily bright at the near-IR.
The detached PPN shells generally have warm 
($\sim 100 - 200$ K) dust grains, and the IR excess
tends to peak at the redward of 20 $\um$.
The near-IR region of the PPN spectrum corresponds to 
the ``trough'' of the double-peaked SED structure.
It is, however, significantly important to determine 
the exact SED shapes of PPNs from optical to far-IR
by filling the gap at the near-IR.

Recent studies of the PPN shell morphology have shown 
that there are two types of PPNs (SOLE-toroidal and 
DUPLEX-core/elliptical types), which are manifestations
of distinct {\sl physical} conditions in the PPN shells
\citep{ueta00,meixner02}.
The optical depth of the shell is the origin of the 
morphological dichotomy among PPNs, and this dichotomy 
can also be seen in the SEDs.
The SEDs are either fully double-peaked or dominated by 
the IR excess: the central star suffers from more 
extinction when surrounded by an optically thicker 
dust shell.
Near-IR photometry helps to classify these PPNs as either 
DUPLEX or SOLE type 
(see Figures 6, 7, and 8 in \citealt{ueta00}).

Moreover, knowledge of the SED shape is often important
in modeling the PPN shells through radiative transfer
calculations.
The near-IR photometric data define the spectral region
corresponding to the redward slope of the optical peak
to the trough of the double-peaked SED of PPNs.
This part of the SED is critically affected by the
basic parameters of the central star ($\teff$, $\lstar$,
and distance) which, of course, would determine the
energy budget for dust heating in the shell
\citep{ueta01a,ueta01b,meixner02}.
Near-IR emission can also arise from some other means.
For example, part of the near-IR excess emission 
from a PPN, IRAS 22272+5435, has been attributed to 
a transiently heated hot dust component \citep{szczerba97}, 
which may have been ejected by a sudden post-AGB mass 
ejection \citep{hrivnak94,ueta01b}.

Motivated by the importance of near-IR photometric data
for PPNs, we have undertaken a series of observations
aiming to obtain accurate near-IR photometry of PPN 
candidates by means of near-IR imaging.
Previous large scale photometric surveys of PPNs used 
photometers with large beam sizes
(8$\arcsec$ to 16$\arcsec$ by \citealt{veen89} and
15$\arcsec$ by \citealt{manchado89}, \citealt{garcialario90}, 
and \citealt{garcialario97}) and sources in the crowded
field may have been misidentified and/or their photometry
inaccurately determined by confusion.
This problem has already been addressed in recent near-IR 
imaging photometric observations for southern post-AGB 
stars \citep{vandesteene00}.
We have also looked for a possible near-IR variability
for our target sources by observing the same sources
at different epochs and also by comparing our results
with past measurements.

In this paper, we present near-IR magnitudes in standard
$J$, $H$, and $\kp$ bands for mostly northern PPN candidates.
Most of the target sources are selected from the list
given by \citet{meixner99} for which we have obtained 
mid-IR photometric data and images.

\section{Observations and Data Reduction\label{obs}}

Near-IR photometric data were taken with the 40 inch
telescope at the Mt.\ Laguna Observatory\footnote{%
Mt.\ Laguna Observatory is jointly 
operated by University of Illinois at Urbana-Champaign
and San Diego State University.}
in Southern California
using the Near-IR Imager (NIRIM, \citealt{nirim}).
NIRIM has a Rockwell NICMOS3 Hg:Cd:Te 256 $\times$ 256
array detector.
The pixel scales were set to $0\farcs5$ pixel$^{-1}$, 
$1\farcs0$ pixel$^{-1}$, and $2\farcs0$ pixel$^{-1}$.
While the $1\farcs0$ pixel$^{-1}$ scale was used by
default, we adjusted the plate scale depending on the
brightness of the sources not to saturate the pixels
or to detect the source with a reasonable amount of 
exposure time.

The observations were performed during three different 
runs in 1997 July, 1998 June, and 1999 November.
The log of observations is given in Table \ref{obslog},
showing the pixel scale used, weather conditions, and 
seeing of the night.
Weather conditions were excellent during the 1997 and 
1998 runs, but somewhat chaotic during the 1999 run.
For the ``Mostly Clear'' and ``Clear'' nights during 
the 1999 run, scattered cirrus clouds were seen at
low elevation and humidity was slightly higher ($\sim 40 \%$) 
than usual ($\sim 20 \%$) at the observatory.
Seeing values listed are the average of the FWHMs of
the standard stars at the $J$ band when we checked the
focus of the telescope several times during the night.

In a typical run, we would observe a half dozen 
to a dozen target sources when they are at low 
airmasses ($\ale 1.5$).
We would also observe two to three standard stars
at least twice during the night at low ($\sim 1$) 
and high ($\sim 2$) airmasses to perform an airmass
correction (see below).
Table \ref{standard} shows a list of standard stars
we used during our observing runs.
Data were taken by shifting the telescope with a 
9-point grid dithering pattern.
Each exposure was flat-fielded to eliminate large 
pixel-to-pixel sensitivity variations in the detector 
array. 
Then, the flat-fielded dithered frames were sky-subtracted 
before co-added into a single image.
Flats were created for each waveband by exposing the
twilight sky. 
We constructed sky emission maps by taking a median 
of all dithered frames after unusually low and high 
pixels were masked out.
Our reduction method generally follows the method 
described in \cite{mclean96}.

The correct identification of the target sources is
of prime importance in our observations.
The near-IR sky does not necessarily look the same 
as the optical and mid-IR sky, and this may lead to a 
misidentification of the target sources.
Therefore, we identified the near-IR counterparts
of the target sources in the following way.
First, using the Digital Sky Survey\footnote{%
The Digitized Sky Survey was produced at 
the Space Telescope Science Institute under U.S. 
Government grant NAG W-2166.} (DSS),
we generated optical finding charts of the 
sources centered at their most precise coordinates
available in the literature.
Most reliable coordinates ($\ale 1\arcsec$ accuracy) 
include the values determined by the {\sl Hubble Space 
Telescope} observations and observations of bright 
stars (such as DSS).
For those without any confirmed optical counterparts,
we used coordinates determined by past mid-IR 
observations such as our own (\citealt{meixner99}; 
typically $\ale 5\arcsec$ accuracy) and {\iras} observations
(typically $\ale 10\arcsec$ accuracy).
Then, we took an $I$ band image of a source
and compared it to the optical finding chart of
the source in order to pinpoint the location of the 
near-IR counterpart more precisely.
Our improved identification for many of the
sources comes from translating the source's
optical or mid-IR coordinates into the near-IR 
coordinates via using the $I$ band image.
The accuracy of identification is, however, limited 
by the correctness of the coordinates used,
and this we indicate the basis of the coordinates 
presented in Table \ref{photom}.

The photometry was done using the aperture photometry
package, {\sl apphot}, in IRAF.
Once the target is identified, a polygon aperture is
defined around the star to measure the analog-to-digital
unit (ADU) count within the aperture.
Even though the images are already sky-subtracted,
residual sky counts can still be measured from a sky 
annulus defined surrounding the aperture or sky
boxes in the blank regions of the image.
If the residual sky counts are higher than 5 $\sigma$,
we subtracted the total residual sky counts in the
aperture polygon from the ADU counts of the source.
However, we rarely subtracted the residual sky 
at this stage.

In order to convert the ADU counts to magnitudes, 
we have attempted to determine the zero magnitude 
and the airmass correction factor for each band
for each night. 
Raw ADU counts of the standard stars are converted 
to raw magnitudes.
These raw magnitudes are plotted against airmasses.
Ideally, there should be a linear relationship between 
the raw magnitudes and airmasses, and the airmass
correction factor is obtained as a slope of the 
line fit to the data.
Then, the correction factor can be used to obtain 
the airmass-corrected magnitudes of the standard stars.
Given the magnitudes of the standard stars,
we derive the zero magnitudes of the instrument 
for each band for a given night.

However, the raw magnitude versus airmass plot 
generated from the 1997 and 1998 data sets have 
failed to yield a linear relationship between the two: 
the data points are scattered all over the plot.
The data points were scattered mainly because 
there was only a small difference between low 
and high airmass values, especially for the 
1997 data set.
We were unable to derive the airmass correction
factors for these data sets, and thus, did not 
apply any airmass correction except for the 1999
data set.
This introduced relatively large uncertainties 
in our determination of the zero magnitudes and 
propagated into the resulting photometric values.
Furthermore, the precision of stellar photometry
may be affected by undersampling the 
point-spread-function (PSF).
As clearly demonstrated by \citet{lauer99},
intrapixel sensitivity variations of NICMOS3 
arrays can be as high as 0.39 mag depending on 
where the peak of the target source lands on
the array.
Although dithering would reduce the effect, the 
undersampling of the PSF may contribute to the 
uncertainties in the 1997 and some of the 1998 
data because the PSF was typically relatively 
smaller with respect to the pixel scale.
Since the seeing was generally worse in the 1999 
run than the previous runs, the 1999 data seem to
have suffered less from the intrapixel sensitivity 
variations.
Overall, the uncertainties of the measurements
for the 1999 run are less than those for the
1997 and 1998 runs.

Thus, our results have relatively large
uncertainties ($0.1 - 0.3$ mag) due to
these limitations of measurements.
The precision of our observations is adequate for
fixing the SED, but may not be so for monitoring 
small photometric variabilities in the near-IR.
Our results, nevertheless, improve upon previously
published values because our identifications based 
on imaging are better-suited to identify sources than  
observations with a large aperture photometer, which 
are confusion-limited especially in a crowded field.
The influence of airmass in our photometric results 
is considered to be relatively small, since
the target sources were observed generally at
low airmass ($\ale 1.5$).
A typical atmospheric extinction in the $J$ band
at Mauna Kea is 0.12 mag airmass$^{-1}$ \citep{krisciunas87}.
At Mt.\ Laguna, this value is expected to be larger.
Thus, neglecting the airmass correction for low
airmass targets can lead to uncertainties of
about 0.05 mag.
This is still small relative to the uncertainties
associated with determining the zero magnitudes.

\begin{figure}[p]
\plotone{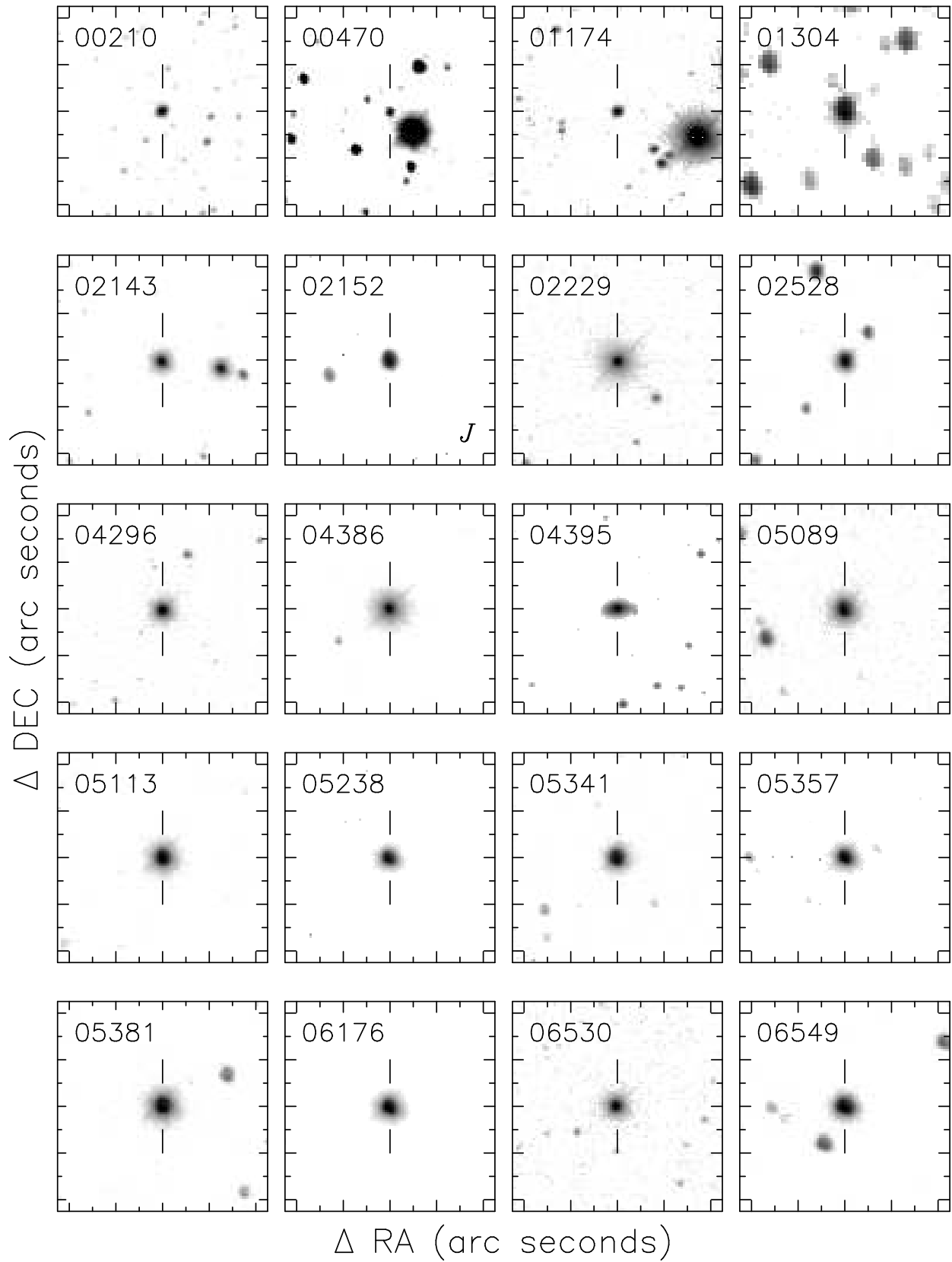}
\figcaption{\label{fc}%
$H$ band finding charts for our target sources,
except for some cases in which a $J$ band chart
is shown (indicated by $J$ at the lower right
in the frame).
The field size is $90\arcsec \times 90\arcsec$
with $10\arcsec$ tickmarks shown. 
N is up and E to the left.  
The object name is indicated on the top right 
corner of each frame.}
\end{figure}

The resulting $JHK^{\prime}$ magnitudes of our
target sources are listed in Table \ref{photom},
along with references to the previous near-IR 
photometries with which we compared our
results.
Also presented in Figure \ref{fc} are $H$ band 
finding charts.
Each frame shows a $90\arcsec \times 90\arcsec$
field centered at the target source.
When the $H$ band image is not available, the 
chart shows the $J$ band image
(indicated by ``J'' in the frame). 
Some stars are too bright to image even at the
0\farcs5 pixel$^{-1}$ scale.
In such cases we have intentionally defocused
the optics not to saturate the detector arrays.
We did not include finder charts for those 
bright, defocused objects, but they are bright 
enough to be easily identified when observed.

\section{Discussion}

\subsection{Misidentification, Variability, and
Extendedness}

As one immediately sees from the finding charts,
many of our target sources turned out to be 
located in a fairly crowded field.
Thus, caution must be exercised when quoting 
previous photometric measurements done using a 
photometer with a large beam.
As have been pointed out and strongly demonstrated
by \citet{vandesteene00} for southern post-AGB
objects, we have discovered 10 cases
in which the source is likely to have been 
misidentified or previous measurements have been 
confused with neighboring objects (see the next
section for details).

We have compared our results with past measurements 
to look for a possible variability, even though our
limited precision may not be sufficient to decisively
determine variabilities at a few tenth of a magnitude.
In doing so, 
one needs to exercise caution because the photometric 
system used may not be the same in different observations. 
Transformation among photometric systems generally 
yielded at most 0.05 mag difference, which is still 
small relative to our uncertainty.
Nevertheless, we have converted the data in the 
literature to the CIT/ELIAS photometric system when 
comparing the past data with our measurements.
The comparison has yielded nearly two dozen cases 
where present and past observations indicate an 
appreciable amount of variation in the near-IR magnitudes, 
including quite a few objects with a previously 
unreported variability.
Follow-up near-IR photometric monitoring observations 
are necessary to determine and characterize their 
variability.

Not all sources are point sources, and some
are so extended that even a 40 inch telescope
can resolve the structure.
Thus, we have also tried to determine if 
any of the targets are extended.
We did not find any obviously extended sources
except for the ones that are known previously
(04395$+$3601, 07399$-$1435, 09371$+$1212,
09452$+$1330, and 17028$-$1004).
There are, however, marginally extended sources
(07134$+$1005, 07284$-$0940, and 08187$-$1905)
and it is worth following up on their 
morphologies using larger aperture telescopes.

\subsection{Notes on Individual Sources}

{\iras} 01174+6110 -- Our magnitudes are about 2 mag
lower than the previously reported values \citep{manchado89}. 
There is a brighter star located about 35$\arcsec$ west 
of the target.
It is conceivable that this bright star may have been 
misidentified, since 60$\arcsec$ wide raster scans with 
a limiting magnitude of 10.5 (at $K$ band) might have 
missed this source ($K = 11.6$).
Otherwise, this source has experienced a significant 
brightness variation.

{\iras} 01304$+$6211 -- Our magnitudes are 1 to 2 mag 
lower than the previously values \citep{jones90,sun91,xiong94}.
Although this source is known as an irregular variable 
\citep{jones90}, confusion-limited past observations
(with a 20$\arcsec$ beam for \citealt{jones90} observations,
for example) may have been partially contaminated by 
the neighboring sources.

{\iras} 02143$+$5852 -- $H$ and $\kp$ magnitudes seem 
to show slight variation compared with the past
observations \citep{garcialario90,fujii02}.

{\iras} 02152$+$2822 -- This source is a known near-IR 
variable \citep{whitelock94}.
Their sporadic $J$ measurements are about 2 mag
dimmer than our value.

{\iras} 02229$+$6208 -- Compared with the previous 
observations by \citet{hrivnak99}, $J$ magnitude 
has become slightly dimmer while $H$ and $\kp$
magnitudes have become roughly a magnitude brighter.
This may be due to recently formed warm dust grains
in the system.

{\iras} 02528$+$4350 -- This source has recently been 
identified as a galaxy (see \citet{fujii02} and reference
therein).

{\iras} 04296$+$3429 -- There seems to be variation 
on the level of 0.2 to 0.4 mag among observations.

{\iras} 04386$+$5722 -- $H$ and $\kp$ brightnesses 
have increased since previously observed by 
\citet{blommaert93}.

{\iras} 04395$+$3601 -- There seems to be variation
up to 2 mag.
However, higher $J$ and $H$ magnitudes reported by
\citet{ueta01c} are at least partly due to the new 
emission structures discovered in their deep imaging.

{\iras} 05089$+$0459 -- Our magnitudes are in agreement 
with those recently measured by \citet{fujii02}, while 
\citet{garcialario90} reported values about 2 mag 
brighter.

{\iras} 05113$+$1347 -- Our measurements are in agreement
with those reported by \citet{garcialario97} and
\citet{fujii02}, while \citet{garcialario90} reported 
about 1 to 2 mag brighter values.

{\iras} 05341$+$1347 -- There seems to be variation 
on the level of 0.2 mag at $J$ and $\kp$ and perhaps 
larger at $H$ among observations.

{\iras} 06549$-$2330 -- \citet{garcialario97} reported
slightly brighter magnitudes than ours.
Their measurements may have been contaminated by the 
nearby near-IR bright star located about 15$\arcsec$ SE.

{\iras} 07399$-$1435 -- This source is extended and 
its brightness variation compared with the past 
observations is well within the level previously 
reported \citep{kastner92}.

{\iras} 07430$+$1115 -- There seems to be variation 
on the level of 0.2 mag among observations.

{\iras} 08005$-$2356 -- There seems to be variation 
on the level of 0.2 mag among observations.

{\iras} 08187$-$1905 -- This known variable has varied its
near-IR magnitude considerably (0.6 mag at $J$, 1 mag at 
$H$, and 1.5 mag at $\kp$) since observed last time nearly
a decade ago \citep{garcialario90}.
The object seems to be extended (FWHM = 3\farcs5).

{\iras} 16342$-$3814 -- Our results are consistent with
the values reported by \citet{veen89}.
Observations by \citet{fouque92} and \citet{garcialario97} 
reported 2 - 3 mag brighter values.
The latter two studies seem to have misidentified the 
source with the nearby brighter star about 10$\arcsec$ 
NE, since our measurements of this nearby source are 
consistent with their reported photometric values.

{\iras} 16559$-$2957 -- There seems to be variation 
on the level of 0.2 - 0.4 mag among observations.

{\iras} 17028$-$1004 -- The brighter magnitudes previously
obtained by \citet{fouque92} and \citet{phillips94} seem 
to have been 
affected by their use of a large-beam photometer and the 
presence of a nearby star roughly 5$\arcsec$ W.

{\iras} 17436$+$5003 -- This known variable varied its
$J$ and $\kp$ magnitudes in the past while keeping
its $H$ magnitude stable.

{\iras} 17441$-$2411 -- There seems to be variation 
on the level of 0.2 - 0.4 mag among observations.

{\iras} 17534$+$2603 -- Our near-IR magnitudes are
1.5 - 2 mag brighter compared with the results
obtained 25 years ago \citep{hn74}.

{\iras} 18071$-$1727 -- The previous observations 
by \citet{veen89} may have been affected by the use of 
a large aperture photometer in this crowded field.

{\iras} 18095$+$2704 -- There seems to be variation 
on the level of 0.3 mag among observations.

{\iras} 18576$+$0341 -- This object is a near-IR 
variable \citep{ueta01a,pasquali02} and is now 
recognized as a luminous blue variable candidate.

{\iras} 19306$+$1407 -- Compared with the previous 
observations by \citet{garcialario97}, $J$ magnitude 
has become slightly dimmer while $H$ and $\kp$
magnitudes have become slightly brighter.
This may be due to recently formed warm dust grains
in the system.

{\iras} 19356$+$0754 -- Only $\kp$ magnitude shows
inconsistency between the present and previous
observations: \citet{garcialario97} reported nearly 
2 mag brighter value.
This is likely caused by confusion with the neighboring 
star 15$\arcsec$ off to the East, which significantly 
brightens in the $\kp$ band.

{\iras} 19477$+$2401 -- \citet{garcialario97} reported
almost 1 to 2 mag brighter values.
The field is rather crowded and their photometric
values are likely to have been suffered from confusion.

{\iras} 20000$+$3239 -- There seems to be variation 
on the level of 0.2 - 0.4 mag among past observations
at $J$ and $\kp$.

{\iras} 20028$+$3910 -- \citet{manchado89} did not detect
the source in their survey with about 10.5 mag detection limit.
Our observations indicate that this source has varied
its brightnesses about 0.5 mag.

{\iras} 20077$-$0625 -- Compared with the previous
observations by \citet{fouque92}, our values are 
roughly 1 mag brighter.

{\iras} 20144$+$4656 -- The nearby bright source roughly 
12$\arcsec$ W seems to have contaminated the previous 
observations \citep{garcialario97} that have yielded 
brighter magnitudes.

{\iras} 20462$+$3416 -- This source is a known variable 
\citep{arkhipova01} and has brighter $H$ and $\kp$ 
magnitudes compared with the previous observations 
\citep{garcialario97}. 

{\iras} 20572$+$4919 -- There seems to be variation 
on the level of 0.3 - 0.6 mag among observations.
\citet{arkhipova00} suggests its optical variability 
is due to a variable stellar wind.
The observed near-IR variability may be related to 
the dust formation in the variable wind.

{\iras} 21289$+$5815 -- There seems to be variation 
on the level of 0.5 mag among observations.

{\iras} 22142$+$5206 -- The previous observations 
by \citet{manchado89} may have been affected by the use of 
a large aperture photometer in this crowded field.

{\iras} 22223$+$4327 -- There seems to be variation 
on the level of 0.3 mag among observations. 
Especially, both present and past \citep{garcialario97} 
two-epoch observations show similar variation,
implying that the near-IR variation is real. 

{\iras} 22272$+$5435 -- There seems to be variation
on the level of 0.3 to 0.4 at $J$ and $\kp$ while
it is less than 0.1 mag at $H$.

{\iras} 22480$+$6002 -- There are discrepancies of 
about 1 mag between our values and previous 
measurements \citep{humphreys74}.

{\iras} 22574$+$6609 -- This source seems to be a variable;
its magnitudes are consistent with past measurements, but 
its $J$ magnitude may be dimmer about 0.5 mag.

{\iras} 23304$+$6147 -- There seems to be variation
on the level of 0.6 - 0.8 mag among observations.

\section{Conclusions}

We have observed 78 objects including PPN candidates
and others and measured their $JHK^{\prime}$
photometric magnitudes by means of imaging.
$JHK^{\prime}$ magnitudes are presented in a tabular 
format together with $H$ band finding charts when available.
Compared with the technique of aperture photometry 
using a photometer with a large beam size, imaging
photometry yields more accurate results.
However, the precision of our measurements are somewhat
limited due to relatively large uncertainties
originated from the zero magnitude determination and 
the intrapixel sensitivity variation of the detector 
arrays.
We have discovered 10 cases (more than 10\%) in 
which previous observations have either misidentified 
the source or been contaminated by neighboring objects 
in a crowded field.
Therefore, caution must be exercised when one
quotes photometric measurements done using a 
photometer with a large beam.
We also have found quite a few sources indicating
a variability, including ones previously unreported,
which needs to be confirmed by follow-up monitoring
observations.
Extended structures have been recognized in some
sources and marginal extension is seen in some
objects from which no high-resolution imaging
has been performed previously.
Follow-up high-resolution imaging of those
objects are also recommended.

\acknowledgments
The authors are supported by NSF CAREER Award 
grant, AST 97-33697.
We acknowledge assistance from Anthony 
Marcotte and Philip Swanson in data reduction.
The anonymous referee is thanked for thorough
review of the manuscript and valuable comments.

\clearpage

\begin{deluxetable}{ccccccc} 
\tablecolumns{5} 
\tablewidth{0pc} 
\tablecaption{\label{obslog}%
Observation Log for Near-IR Photometry for PPN candidates} 
\tablehead{%
\colhead{Run} & 
\colhead{Date} & 
\colhead{Pixel Scale} & 
\colhead{Conditions} & 
\colhead{Seeing}}

\startdata 
1 & 
1997 Jul 23 & 
1\farcs0 & 
Photometric & 
1\farcs6 \\

  & 
\phm{1997} Jul 24 & 
1\farcs0 & 
Photometric & 
1\farcs9 \\

  & 
\phm{1997} Jul 25 & 
1\farcs0 & 
Photometric &
2\farcs0 \\

  & 
\phm{1997} Jul 26 & 
0\farcs5 & 
Clear & 
1\farcs5 \\

2 & 
1998 Jun 19 &    
\nodata & 
Fog & 
\nodata \\

  & 
\phm{1998} Jun 20 & 
1\farcs0 & 
Photometric & 
2\farcs2 \\

  & 
\phm{1998} Jun 21 & 
0\farcs5 & 
Clear & 
2\farcs7 \\

  & 
\phm{1998} Jun 22 & 
1\farcs0 & 
Photometric & 
2\farcs0 \\

  & 
\phm{1998} Jun 23 & 
2\farcs0 & 
Photometric & 
1\farcs9 \\

  & 
\phm{1998} Jun 24 & 
1\farcs0 & 
Photometric & 
2\farcs0 \\

  & 
\phm{1998} Jun 25 & 
1\farcs0 & 
Photometric & 
2\farcs2 \\

  & 
\phm{1998} Jun 26 & 
1\farcs0 & 
Clear & 
3\farcs0 \\

  & 
\phm{1998} Jun 27 & 
2\farcs0 & 
Clear & 
3\farcs5 \\

3  & 
1999 Nov 12 & 
1\farcs0 & 
Photometric & 
2\farcs1 \\

  & 
\phm{1999} Nov 13 & 
1\farcs0 & 
Mostly Clear & 
2\farcs0 \\

  & 
\phm{1999} Nov 14 & 
1\farcs0 & 
Mostly Clear & 
2\farcs2 \\

  & 
\phm{1999} Nov 15 & 
1\farcs0 & 
Clear & 
2\farcs3 \\

  & 
\phm{1999} Nov 16 & 
0\farcs5 & 
Photometric & 
2\farcs1 \\

  & 
\phm{1999} Nov 17 & 
\nodata & 
Cloudy & 
\nodata \\

  & 
\phm{1999} Nov 18 & 
2\farcs0 & 
Photometric & 
2\farcs2 

\enddata
\end{deluxetable}

\begin{deluxetable}{lccccccc} 
\tabletypesize{\small}
\tablecolumns{8} 
\tablewidth{0pc} 
\tablecaption{\label{standard}%
List of the Standard Stars}
\tablehead{%
\colhead{} &
\multicolumn{2}{c}{Coordinates (J2000.0)} & 
\colhead{} &
\multicolumn{3}{c}{Observed Dates} & 
\colhead{} \\
\cline{2-3}
\cline{5-7}
\colhead{Name} & 
\colhead{RA} & 
\colhead{DEC} & 
\colhead{} &
\colhead{1997 Jul} &
\colhead{1998 Jun} &
\colhead{1999 Nov} &
\colhead{Ref.}}

\startdata
HD 225023 & 00 02 46.0 & $+$35 48 56 & & \nodata & \nodata & 13, 17 & 1 \\
HD 1160   & 00 15 57.3 & $+$04 15 04 & & \nodata & \nodata & 15 & 1 \\
HD 3029   & 00 33 39.5 & $+$20 26 02 & & \nodata & \nodata & 13 & 1 \\
HD 18881  & 03 03 31.9 & $+$38 24 36 & & \nodata & \nodata & 13, 14, 16, 18 & 1 \\
HD 22686  & 03 38 55.1 & $+$02 45 49 & & \nodata & \nodata & 12, 16 & 1 \\
FS 11     & 04 52 58.9 & $-$00 14 42 & & \nodata & \nodata & 14 & 2 \\
HD 40335  & 05 58 13.5 & $+$01 51 23 & & \nodata & \nodata & 15 & 1 \\
FS 23     & 13 41 43.6 & $+$28 29 50 & & \nodata & 23 & \nodata & 2 \\
HD 129655 & 14 43 46.4 & $-$02 30 20 & & \nodata & 22, 24 & \nodata & 1 \\
HD 136754 & 15 21 34.5 & $+$24 20 36 & & \nodata & 21 & \nodata & 1 \\
S-R 3     & 16 23 07.7 & $-$24 27 26 & & \nodata & 23 & \nodata & 1 \\
HD 162208 & 17 47 58.6 & $+$39 58 51 & & 26 & 25 & \nodata & 1 \\
HD 161903 & 17 48 19.2 & $-$01 48 30 & & 23 & 20, 21, 24 & \nodata & 1 \\
FS 35     & 18 27 13.5 & $+$04 03 09 & & 23, 24 & 27 & \nodata & 2 \\
GL 748    & 19 12 14.6 & $+$02 53 11 & & \nodata & 20, 21, 22, 24, 26 & \nodata & 1 \\ 
GL 811.1  & 20 56 46.6 & $-$10 26 55 & & 26 & 23, 27 & \nodata & 1 \\
HD 203856 & 21 23 35.5 & $+$40 01 07 & & 23, 24, 25 & 26 & 12, 14, 16, 18 & 1 \\
FS 31     & 23 12 21.6 & $+$10 47 04 & & 25 & \nodata & \nodata & 2 
\enddata 
\tablerefs{1. \citet{elias82}, 2. \citet{casali92}}
\end{deluxetable}

\begin{deluxetable}{lccccccccc} 
\tabletypesize{\scriptsize}
\tablecolumns{8} 
\tablewidth{0pc} 
\tablecaption{\label{photom}%
Near-Infraed Photometry of PPN Candidates} 
\tablehead{%
\colhead{Source} &
\multicolumn{3}{c}{Coordinates (J2000.0)} & 
\colhead{} &
\multicolumn{3}{c}{Magnitudes} & 
\colhead{Comparison} \\
\cline{2-4}
\cline{6-8}
\colhead{(Other Name)} &
\colhead{RA} & 
\colhead{DEC} &
\colhead{Ref.\tablenotemark{a}} &
\colhead{Date}   & 
\colhead{$J$}    & 
\colhead{$H$}    & 
\colhead{$K^{\prime}$} & 
\colhead{Ref.}}

\startdata 
00210$+$6221 & 
00 23 51.1 & 
$+$62 38 16 &
MIR & 
1999 Nov 12 & 
12.90$\pm$0.06 & 
11.93$\pm$0.05 & 
11.76$\pm$0.07 &
\nodata \\

00470$+$6429 & 
00 50 07.2 & 
$+$64 45 46 & 
MIR & 
1999 Nov 13 &
14.77$\pm$0.07 & 
14.37$\pm$0.10 & 
14.83$\pm$0.22 &
\nodata \\

01174$+$6110 & 
01 20 44.3 & 
$+$61 26 21 &
MIR & 
1999 Nov 12 & 
14.77$\pm$0.07 & 
13.03$\pm$0.05 & 
11.57$\pm$0.07 &
12 \\

01304$+$6211 & 
01 33 51.2 & 
$+$62 26 53 & 
MIR & 
1999 Nov 18 & 
\nodata & 
10.84$\pm$0.07 & 
$\phn$6.85$\pm$0.34 &
19, 21, 31 \\

02143$+$5852 & 
02 17 57.9 & 
$+$59 05 51 & 
MIR & 
1999 Nov 13 & 
10.59$\pm$0.06 & 
$\phn$9.50$\pm$0.04 &
$\phn$8.46$\pm$0.07 &
37, 40 \\

02152$+$2822 & 
02 18 06.6 & 
$+$28 36 48 & 
MIR & 
1999 Nov 15 & 
10.51$\pm$0.17 & 
\nodata & 
\nodata &
30 \\

02229$+$6208 & 
02 26 41.9 & 
$+$62 21 22 &
HST &  
1999 Nov 12 & 
$\phn$6.83$\pm$0.06 &
$\phn$4.65$\pm$0.08 & 
$\phn$4.20$\pm$0.12 &
38 \\

02528$+$4350 & 
02 56 11.3 & 
$+$44 02 51 &
IRAS &  
1999 Nov 13 & 
10.12$\pm$0.06 & 
$\phn$9.82$\pm$0.04 & 
$\phn$9.42$\pm$0.07 &
12, 37, 40 \\

04296$+$3429 & 
04 32 57.0 & 
$+$34 36 13 & 
HST & 
1999 Nov 12 & 
$\phn$9.64$\pm$0.06 & 
$\phn$8.75$\pm$0.04 & 
$\phn$8.29$\pm$0.07 &
12, 17, 37, 40 \\

04386$+$5722 & 
04 42 49.0 & 
$+$57 27 47 & 
HST & 
1999 Nov 12 & 
$\phn$6.75$\pm$0.06 & 
$\phn$4.34$\pm$0.06 & 
$\phn$3.95$\pm$0.09 &
25 \\*
 &
 &
 &
 &
1999 Nov 13 & 
$\phn$6.76$\pm$0.06 & 
$\phn$4.36$\pm$0.07 & 
$\phn$3.92$\pm$0.13 & 
 \\

04395$+$3601 & 
04 42 53.6 & 
$+$36 06 54 &
OPT &  
1999 Nov 12 & 
13.04$\pm$0.08 & 
11.29$\pm$0.04 & 
$\phn$8.81$\pm$0.07 &
24, 34, 39 \\
~(GL 618) & \\

05089$+$0459 & 
05 11 36.0 & 
$+$05 03 26 & 
MIR & 
1999 Nov 14 & 
10.27$\pm$0.06 & 
$\phn$9.26$\pm$0.04 & 
$\phn$8.90$\pm$0.07 &
14, 40 \\*
 &
 &
 &
 & 
1999 Nov 15 & 
10.27$\pm$0.15 & 
$\phn$9.27$\pm$0.07 & 
$\phn$8.81$\pm$0.31 & 
 \\

05113$+$1347 & 
05 14 07.8 & 
$+$13 50 28 & 
HST & 
1999 Nov 12 & 
$\phn$8.94$\pm$0.06 & 
$\phn$8.36$\pm$0.04 & 
$\phn$8.11$\pm$0.07 &
14, 33, 37, 40 \\*
 &
 &
 &
 & 
1999 Nov 14 & 
$\phn$9.04$\pm$0.06 & 
$\phn$8.39$\pm$0.04 & 
$\phn$8.17$\pm$0.07 &
 \\

05238$-$0626 & 
05 26 19.8 & 
$-$06 23 57 & 
OPT & 
1999 Nov 15 & 
$\phn$9.66$\pm$0.15 & 
$\phn$9.35$\pm$0.07 & 
$\phn$9.14$\pm$0.31 &
14, 37, 40 \\

05341$+$0852 & 
05 36 55.0 & 
$+$08 54 08 & 
HST & 
1999 Nov 14 & 
10.09$\pm$0.06 & 
$\phn$9.45$\pm$0.04 & 
$\phn$9.17$\pm$0.07 &
12, 15, 37, 38, 40 \\

05357$-$0217 & 
05 38 14.1 & 
$-$02 15 60 & 
OPT & 
1999 Nov 15 & 
$\phn$9.47$\pm$0.15 & 
$\phn$9.14$\pm$0.07 & 
$\phn$8.94$\pm$0.31 &
14 \\

05381$+$1012 & 
05 40 56.7 & 
$+$10 14 25 & 
IRAS & 
1999 Nov 14 & 
$\phn$8.86$\pm$0.06 & 
$\phn$8.36$\pm$0.04 & 
$\phn$8.23$\pm$0.07 &
12 \\*
 &
 &
 &
 & 
1999 Nov 15 & 
$\phn$8.83$\pm$0.15 & 
$\phn$8.40$\pm$0.07 & 
$\phn$8.17$\pm$0.31 &
 \\

06176$-$1036 & 
06 19 58.2 & 
$-$10 38 15 & 
OPT & 
1999 Nov 15 & 
$\phn$6.59$\pm$0.15 & 
$\phn$5.01$\pm$0.10 & 
\nodata &
22 \\*
~(Red Rectangle) &
 &
 &
 & 
1999 Nov 16 & 
$\phn$6.50$\pm$0.15 & 
$\phn$4.97$\pm$0.08 & 
$\phn$3.98$\pm$0.44 &
 \\

06530$-$0213 & 
06 55 31.8 &
$-$02 17 29 &  
HST & 
1999 Nov 12 & 
$\phn$9.56$\pm$0.06 & 
$\phn$8.86$\pm$0.04 & 
$\phn$8.48$\pm$0.07 &
26, 37 \\

06549$-$2330 & 
06 57 05.3 & 
$-$23 34 18 & 
IRAS & 
1999 Nov 15 & 
$\phn$9.31$\pm$0.15 & 
$\phn$9.07$\pm$0.07 & 
$\phn$8.91$\pm$0.31 &
37 \\

07134$+$1005 & 
07 16 10.3 & 
$+$09 59 48 & 
HST & 
1999 Nov 16 & 
$\phn$6.79$\pm$0.15 & 
$\phn$6.61$\pm$0.07 & 
$\phn$7.03$\pm$0.31 &
11, 12 \\*
~(HD 56216) & \\

07227$-$1320 & 
07 25 03.1 & 
$-$13 26 17 & 
IRAS & 
1999 Nov 15 & 
$\phn$8.83$\pm$0.15 & 
$\phn$7.89$\pm$0.07 & 
$\phn$7.53$\pm$0.31 &
37 \\

07239$-$3325 & 
07 25 52.2 & 
$-$33 31 22 & 
IRAS & 
1999 Nov 15 & 
$\phn$9.08$\pm$0.15 & 
$\phn$8.05$\pm$0.08 & 
$\phn$7.69$\pm$0.33 &
\nodata \\

07284$-$0940 & 
07 30 47.5 & 
$-$09 46 37 & 
OPT & 
1999 Nov 16 & 
$\phn$4.33$\pm$0.16 & 
$\phn$3.90$\pm$0.11 & 
$\phn$4.16$\pm$0.45 &
\nodata \\*
~(U Mon) & \\

07331$+$0021 & 
07 35 41.2 & 
$+$00 14 58 & 
OPT & 
1999 Nov 16 & 
$\phn$5.92$\pm$0.15 & 
$\phn$5.46$\pm$0.07 & 
$\phn$5.63$\pm$0.32 &
14 \\

07399$-$1435 & 
07 42 16.8 & 
$-$14 42 52 & 
IRAS & 
1999 Nov 15 & 
$\phn$8.29$\pm$0.15 & 
$\phn$6.92$\pm$0.07 & 
$\phn$5.78$\pm$0.32 &
23, 35, 37 \\*
~(OH 231.8) & \\

07430$+$1115 & 
07 45 51.4 & 
$+$11 08 20 & 
HST & 
1999 Nov 12 & 
$\phn$9.03$\pm$0.06 & 
$\phn$8.31$\pm$0.04 & 
$\phn$7.94$\pm$0.07 &
37, 38, 40 \\

07506$-$0345 & 
07 53 06.9 & 
$-$03 53 29 & 
IRAS & 
1999 Nov 14 & 
13.44$\pm$0.11 & 
\nodata & 
\nodata &
14 \\

08005$-$2356 & 
08 02 40.8 & 
$-$24 04 44 & 
HST & 
1999 Nov 15 & 
$\phn$8.02$\pm$0.15 & 
$\phn$6.90$\pm$0.08 & 
$\phn$5.64$\pm$0.42 &
20, 35, 37 \\

08187$-$1905 & 
08 20 57.1 & 
$-$19 15 04 & 
OPT & 
1999 Nov 16 & 
$\phn$7.93$\pm$0.15 & 
$\phn$8.37$\pm$0.07 & 
$\phn$8.79$\pm$0.31 &
14 \\

09371$+$1212 & 
09 39 53.6 & 
$+$11 58 54 &
IRAS & 
1999 Nov 12 & 
$\phn$7.96$\pm$0.06 & 
$\phn$7.41$\pm$0.04 & 
$\phn$7.39$\pm$0.07 &
7, 10 \\*
~(Frosty Leo) \\

09452$+$1330 & 
09 47 57.4 & 
$+$13 16 43 & 
HST & 
1999 Nov 16 & 
$\phn$6.85$\pm$0.15 & 
$\phn$3.64$\pm$0.07 & 
$\phn$0.20$\pm$0.40 &
9 \\*
~(IRC +10216) & \\

10131$+$3049 & 
10 16 02 & 
$+$30 34 19 & 
MIR & 
1999 Nov 16 & 
$\phn$7.06$\pm$0.15 & 
$\phn$4.34$\pm$0.07 & 
$\phn$1.43$\pm$0.31 &
3, 4 \\*
~(CIT 6) & \\

15465$+$2818 & 
15 48 34.4 & 
$+$28 09 25 & 
HST & 
1998 Jun 21 & 
$\phn$5.97$\pm$0.21 & 
$\phn$5.42$\pm$0.17 & 
$\phn$4.81$\pm$0.23 &
1, 36 \\*
~(R CrB) & \\

16342$-$3814 & 
16 37 39.9 & 
$-$38 20 17 & 
HST & 
1998 Jun 20 & 
11.70$\pm$0.23 & 
10.54$\pm$0.19 & 
$\phn$9.64$\pm$0.27 &
13, 22, 37 \\

16559$-$2957 & 
16 59 08.1 & 
$-$30 01 41 & 
MIR & 
1998 Jun 20 & 
11.80$\pm$0.24 & 
10.83$\pm$0.20 & 
$\phn$9.51$\pm$0.25 &
14, 22, 26 \\

17028$-$1004 & 
17 05 37.7 & 
$-$10 08 33 & 
OPT & 
1998 Jun 20 & 
10.62$\pm$0.24 & 
12.75$\pm$0.27 & 
10.90$\pm$0.99 &
22, 29 \\*
~(M 2-9) & \\

17436$+$5003 & 
17 44 55.4 & 
$+$50 02 40 &
HST &  
1997 Jul 26 & 
$\phn$6.22$\pm$0.12 & 
$\phn$5.98$\pm$0.13 & 
$\phn$6.04$\pm$0.17 &
1, 14 \\*
~(HD 161796) & \\

17441$-$2411 & 
17 47 13.5 & 
$-$24 12 50 & 
HST & 
1997 Jul 23 & 
11.12$\pm$0.13 & 
10.14$\pm$0.14 & 
$\phn$9.51$\pm$0.19 &
13, 26, 37 \\

Sakurai's Object & 
17 52 32.6 & 
$-$17 41 08 & 
OPT & 
1998 Jun 25 & 
$\phn$7.41$\pm$0.22 & 
$\phn$6.25$\pm$0.28 & 
$\phn$5.12$\pm$0.44 & 
41 \\*
~(V4334 Sgr) &
 &
 &
 & 
1998 Jun 26 & 
$\phn$7.41$\pm$0.21 & 
$\phn$6.24$\pm$0.18 & 
$\phn$5.23$\pm$0.29 &
 \\

17534$+$2603 & 
17 55 25.2 & 
$+$26 02 60 & 
OPT & 
1998 Jun 21 & 
$\phn$4.87$\pm$0.21 & 
$\phn$4.43$\pm$0.17 & 
$\phn$3.61$\pm$0.23 &
1 \\*
~(89 Her) & \\

18071$-$1727 & 
18 10 05.9 & 
$-$17 26 35 & 
MIR & 
1997 Jul 24 & 
15.57$\pm$0.22 & 
14.78$\pm$0.44 & 
12.99$\pm$0.31 &
13 \\

18095$+$2704 & 
18 11 30.8 & 
$+$27 05 15 & 
HST & 
1998 Jun 22 & 
$\phn$7.27$\pm$0.21 & 
$\phn$6.73$\pm$0.17 & 
$\phn$6.38$\pm$0.23 &
8, 13, 16, 32, 37 \\

18184$-$1302 & 
18 21 15.9 & 
$-$13 01 27 & 
MIR & 
1998 Jun 21 & 
$\phn$9.17$\pm$0.21 & 
$\phn$7.55$\pm$0.17 & 
$\phn$6.87$\pm$0.11 &
5 \\*
~(MWC 922) & \\

18184$-$1623 & 
18 21 19.5 & 
$-$16 22 26 & 
MIR & 
1997 Jul 26 & 
$\phn$5.09$\pm$0.44 & 
$\phn$4.58$\pm$0.51 & 
$\phn$4.40$\pm$0.58 &
12 \\*
~(HD 168625) &
 &
 &
 & 
1998 Jun 21 & 
$\phn$5.19$\pm$0.24 & 
$\phn$4.65$\pm$0.25 & 
$\phn$4.36$\pm$0.40 &
 \\

18276$-$1431 & 
18 30 30.5 & 
$-$14 28 53 & 
MIR & 
1998 Jun 24 & 
11.54$\pm$0.22 & 
10.62$\pm$0.19 & 
$\phn$9.33$\pm$0.24 &
6, 26 \\

18576$+$0341 & 
19 00 11.2 & 
$+$03 45 46 & 
IRAS & 
1999 Nov 14 & 
12.44$\pm$0.07 & 
$\phn$9.32$\pm$0.05 & 
$\phn$7.60$\pm$0.08 &
37, 42 \\

19075$+$0921 & 
19 09 57.1 & 
$+$09 26 52 & 
MIR & 
1998 Jun 20 & 
14.93$\pm$1.31 & 
12.39$\pm$0.35 & 
11.03$\pm$0.29 &
\nodata \\

19192$+$0922 & 
19 21 36.5 & 
$+$09 27 56 & 
MIR & 
1998 Jun 27 & 
$\phn$8.91$\pm$0.21 & 
$\phn$6.56$\pm$0.18 & 
\nodata &
\nodata \\

19244$+$1115 & 
19 26 48.0 & 
$+$11 21 17 & 
OPT & 
1997 Jul 26 & 
$\phn$5.52$\pm$0.13 & 
$\phn$4.52$\pm$0.17 & 
$\phn$3.69$\pm$0.22 &
27 \\*
~(IRC +10420) &
 &
 &
 & 
1998 Jun 21 & 
$\phn$5.59$\pm$0.21 & 
$\phn$4.62$\pm$0.17 & 
$\phn$3.74$\pm$0.24 &
 \\

19306$+$1407 & 
19 32 55.1 & 
$+$14 13 36 & 
IRAS & 
1999 Nov 15 & 
11.42$\pm$0.15 & 
10.80$\pm$0.08 & 
10.35$\pm$0.31 &
37 \\

19327$+$3024 & 
19 34 45.2 & 
$+$30 30 59 & 
OPT & 
1999 Nov 15 & 
$\phn$8.98$\pm$0.15 & 
$\phn$8.80$\pm$0.07 & 
$\phn$7.78$\pm$0.31 &
29 \\*
~(BD+30$^{\circ}$3639) & \\

19356$+$0754 & 
19 38 01.9 & 
$+$08 01 32 & 
IRAS & 
1999 Nov 15 & 
11.57$\pm$0.15 & 
10.89$\pm$0.07 & 
10.59$\pm$0.31 &
37 \\

19386$+$0155 & 
19 41 08.2 & 
$+$02 02 31 & 
MIR & 
1997 Jul 23 & 
$\phn$7.94$\pm$0.12 & 
$\phn$6.96$\pm$0.13 & 
$\phn$6.07$\pm$0.19 &
13, 22 \\

19475$+$3119 & 
19 49 29.6 & 
$+$31 27 16 & 
OPT & 
1999 Nov 15 & 
$\phn$7.87$\pm$0.15 & 
$\phn$7.57$\pm$0.07 & 
$\phn$7.37$\pm$0.34 &
37 \\

19477$+$2401 & 
19 49 54.4 & 
$+$24 08 51 & 
MIR & 
1998 Jun 20 & 
12.77$\pm$0.24 & 
11.50$\pm$0.18 & 
11.20$\pm$0.25 &
37 \\

20000$+$3239 & 
20 01 59.4 & 
$+$32 47 32 & 
MIR & 
1997 Jul 23 & 
$\phn$7.99$\pm$0.13 & 
$\phn$6.98$\pm$0.12 & 
$\phn$6.57$\pm$0.17 &
12, 33 \\*
 &
 &
 &
 & 
1999 Nov 18 & 
$\phn$7.96$\pm$0.15 & 
$\phn$6.97$\pm$0.07 & 
$\phn$6.51$\pm$0.40 &
 \\

20004$+$2955 & 
20 02 27.3 & 
$+$30 04 25 & 
MIR & 
1997 Jul 26 & 
$\phn$4.99$\pm$0.16 & 
$\phn$4.23$\pm$0.16 & 
$\phn$3.97$\pm$0.21 &
11 \\*
 &
 &
 &
 & 
1998 Jun 21 & 
$\phn$5.23$\pm$0.21 & 
$\phn$4.25$\pm$0.17 & 
$\phn$3.91$\pm$0.23 & 
 \\

20028$+$3910 & 
20 04 35.9 & 
$+$39 18 45 & 
HST & 
1997 Jul 25 & 
14.91$\pm$0.25 & 
14.77$\pm$0.42 & 
14.30$\pm$0.75 &
12 \\*
 &
 &
 &
 & 
1999 Nov 13 & 
15.72$\pm$0.11 & 
16.36$\pm$0.11 & 
14.52$\pm$0.27 &
 \\

20042$+$3259 & 
20 06 10.6 & 
$+$33 07 51 &
IRAS &  
1999 Nov 18 & 
13.95$\pm$0.16 & 
13.43$\pm$0.12 & 
13.14$\pm$0.32 &
37 \\
 
20043$+$2653 & 
20 06 22.7 & 
$+$27 02 32 & 
HST & 
1997 Jul 25 & 
\nodata & 
13.78$\pm$0.15 & 
10.32$\pm$0.18 &
12 \\
 
20077$-$0625 & 
20 10 27.4 & 
$-$06 16 16 & 
MIR & 
1998 Jun 24 & 
$\phn$5.43$\pm$0.25 & 
$\phn$3.14$\pm$0.20 & 
\nodata &
22 \\

20136$+$1309 & 
20 16 00.1 & 
$+$13 18 55 & 
IRAS & 
1999 Nov 15 & 
10.40$\pm$0.15 & 
$\phn$9.34$\pm$0.07 & 
$\phn$8.45$\pm$0.31 &
12 \\

20144$+$4656 & 
20 15 58.3 & 
$+$47 05 39 & 
IRAS & 
1999 Nov 18 & 
13.98$\pm$0.22 & 
\nodata & 
\nodata &
37 \\

20461$+$3853 & 
20 48 04.6 & 
$+$39 05 00 & 
IRAS & 
1999 Nov 18 & 
11.45$\pm$0.15 & 
10.33$\pm$0.07 & 
$\phn$9.61$\pm$0.31 &
37 \\

20462$+$3416 & 
20 48 16.6 & 
$+$34 27 25 & 
HST & 
1999 Nov 12 & 
10.43$\pm$0.06 & 
$\phn$9.82$\pm$0.04 & 
$\phn$9.56$\pm$0.07 &
37 \\*
~(V1853 Cyg) & \\

20572$+$4919 & 
20 58 55.7 & 
$+$49 31 12 & 
OPT & 
1999 Nov 14 & 
$\phn$9.09$\pm$0.06 & 
$\phn$8.17$\pm$0.04 & 
$\phn$7.39$\pm$0.07 &
11, 37 \\*
 &
 &
 &
 &
1999 Nov 18 & 
$\phn$8.98$\pm$0.15 & 
$\phn$8.01$\pm$0.07 & 
$\phn$7.13$\pm$0.36 &
 \\

21282$+$5050 & 
21 29 58.5 & 
$+$51 04 00 & 
IRAS & 
1999 Nov 13 & 
11.63$\pm$0.06 & 
10.77$\pm$0.04 & 
$\phn$9.59$\pm$0.07 &
28 \\

21289$+$5815 & 
21 30 23.0 & 
$+$58 28 51 & 
IRAS & 
1999 Nov 18 & 
12.53$\pm$0.16 & 
11.63$\pm$0.10 & 
10.40$\pm$0.31 &
37 \\

21525$+$5643 & 
21 54 15.2 & 
$+$56 57 25 & 
IRAS & 
1999 Nov 18 & 
14.87$\pm$0.17 & 
13.96$\pm$0.22 & 
12.90$\pm$0.33 &
12 \\

22142$+$5206 & 
22 16 10.1 & 
$+$52 21 13 &
HST &  
1997 Jul 24 & 
15.19$\pm$0.15 & 
12.32$\pm$0.12 & 
10.92$\pm$0.18 &
12 \\

22223$+$4327 & 
22 24 31\phd\phn & 
$+$43 43 09 & 
MIR & 
1997 Jul 23 & 
$\phn$7.69$\pm$0.12 & 
$\phn$7.26$\pm$0.12 & 
$\phn$7.13$\pm$0.17 &
33, 37 \\
 &
 &
 &
 & 
1998 Jun 26 & 
$\phn$7.96$\pm$0.21 & 
$\phn$7.52$\pm$0.17 &
$\phn$7.31$\pm$0.23 &
 \\

22272$+$5435 & 
22 29 10.4 & 
$+$54 51 07 & 
HST & 
1999 Nov 16 & 
$\phn$5.43$\pm$0.15 & 
$\phn$4.89$\pm$0.07 & 
$\phn$5.10$\pm$0.32 &
12, 13, 18, 37 \\*
~(HD 235858) & \\

22480$+$6002 & 
22 49 59.1 & 
$+$60 17 55 & 
MIR & 
1997 Jul 26 & 
$\phn$4.99$\pm$0.16 & 
$\phn$4.23$\pm$0.16 & 
$\phn$3.97$\pm$0.21 &
2 \\

22574$+$6609 & 
22 59 18.3 & 
$+$66 25 47 & 
HST & 
1997 Jul 25 & 
14.00$\pm$0.14 & 
13.62$\pm$0.17 & 
13.18$\pm$0.20 &
17 \\
 &
 &
 &
 & 
1999 Nov 14 & 
13.83$\pm$0.08 & 
13.15$\pm$0.11 & 
12.91$\pm$0.13 &
 \\

23166$+$1655 & 
23 19 12.3 & 
$+$17 11 35 & 
MIR & 
1998 Jun 26 & 
10.72$\pm$0.21 & 
10.43$\pm$0.17 & 
10.90$\pm$0.24 &
14 \\

23304$+$6147 & 
23 32 45.0 & 
$+$62 03 49 & 
IRAS & 
1997 Jul 25 & 
$\phn$8.43$\pm$0.12 & 
$\phn$7.70$\pm$0.12 & 
$\phn$7.40$\pm$0.18 &
12, 14, 17, 37, 40 \\
 &
 &
 &
 & 
1998 Jun 27 & 
$\phn$9.06$\pm$0.21 & 
$\phn$8.48$\pm$0.18 & 
$\phn$8.01$\pm$0.25 &
 \\

23321$+$6545 & 
23 34 23.1 & 
$+$66 01 51 & 
HST & 
1998 Jun 27 & 
13.92$\pm$0.21 & 
\nodata & 
12.47$\pm$0.59 &
\nodata
\enddata 

\tablenotetext{a}{Basis for the coordinates: 
HST - {\it HST} observations ($\ale 1\arcsec$ accuracy), 
OPT - optical observations ($\ale 1\arcsec$ accuracy),
MIR - mid-IR observations (\citealt{meixner99}; 
typically $\ale 5\arcsec$ accuracy),
IRAS - {\iras} observations (typically $\ale 10\arcsec$ accuracy)}

\tablerefs{%
1.\ \citet{hn74},
2.\ \citet{humphreys74},
3.\ \citet{strecker74},
4.\ \citet{alksnis75},
5.\ \citet{allen77},
6.\ \citet{lebertre87},
7.\ \citet{hodapp88},
8.\ \citet{hrivnak88},
9.\ \citet{lebertre88},
10.\ \citet{rouan88},
11.\ \citet{hrivnak89},
12.\ \citet{manchado89}, 
13.\ \citet{veen89},
14.\ \citet{garcialario90},
15.\ \citet{geballe90},
16.\ \citet{lawrence90},
17.\ \citet{hrivnak91},
18.\ \citet{hrivnak91b},
19.\ \citet{jones90},
20.\ \citet{slijkhuis91},
21.\ \citet{sun91},
22.\ \citet{fouque92},
23.\ \citet{kastner92},
24.\ \citet{latter92},
25.\ \citet{blommaert93},
26.\ \citet{hu93},
27.\ \citet{jones93},
28.\ \citet{khl93},
29.\ \citet{phillips94},
30.\ \citet{whitelock94},
31.\ \citet{xiong94},
32.\ \citet{kastner95},
33.\ \citet{kwok95},
34.\ \citet{latter95},
35.\ \citet{lepine95},
36.\ \citet{feast97},
37.\ \citet{garcialario97},
38.\ \citet{hrivnak99},
39.\ \citet{ueta01c},
40.\ \citet{fujii02},
41.\ \citet{kamath02},
42.\ \citet{pasquali02}}
\end{deluxetable}

\clearpage

\begin{figure}[p]
\plotone{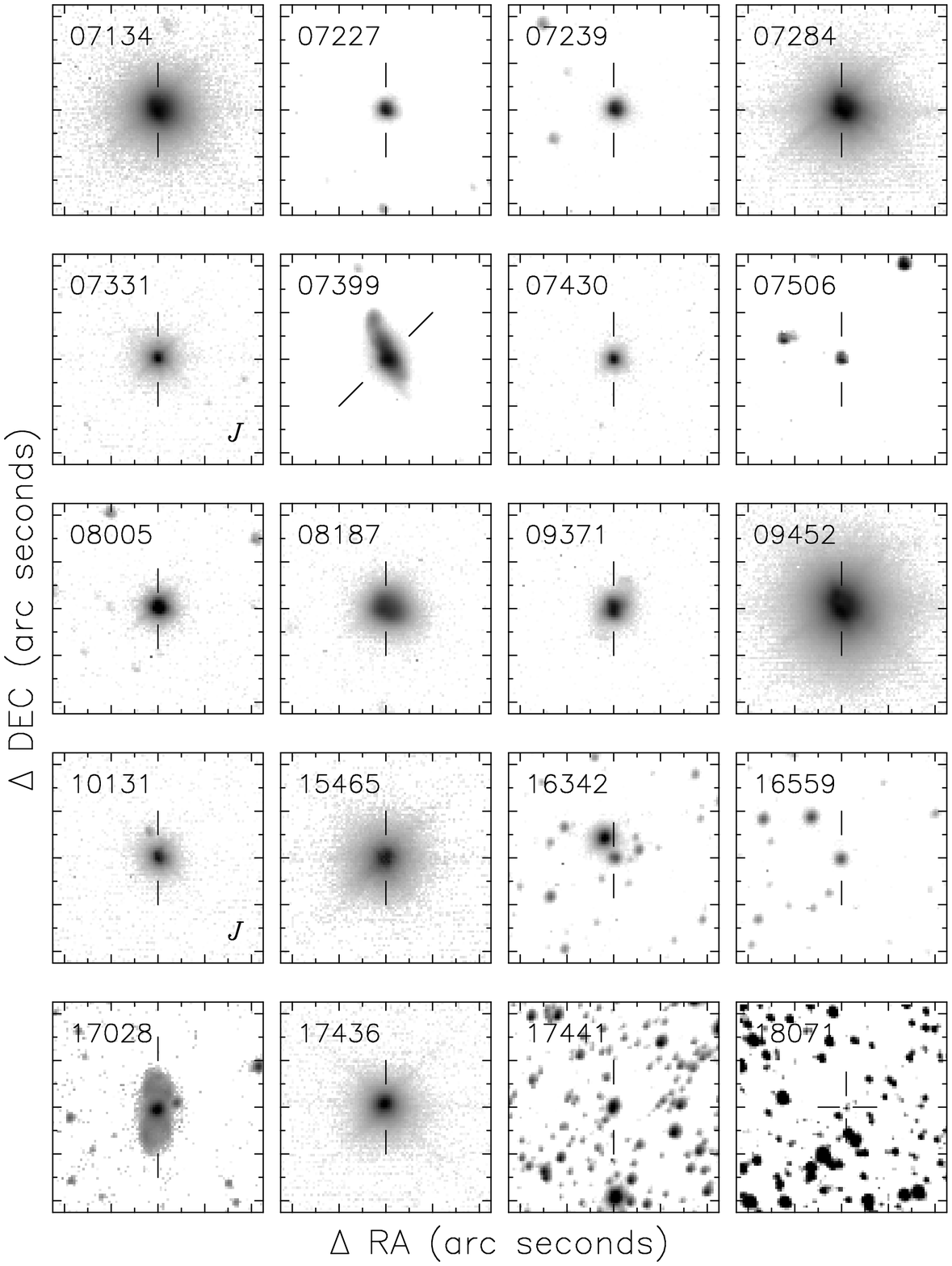}
\figcaption{%
Continued}
\end{figure}

\begin{figure}[p]
\plotone{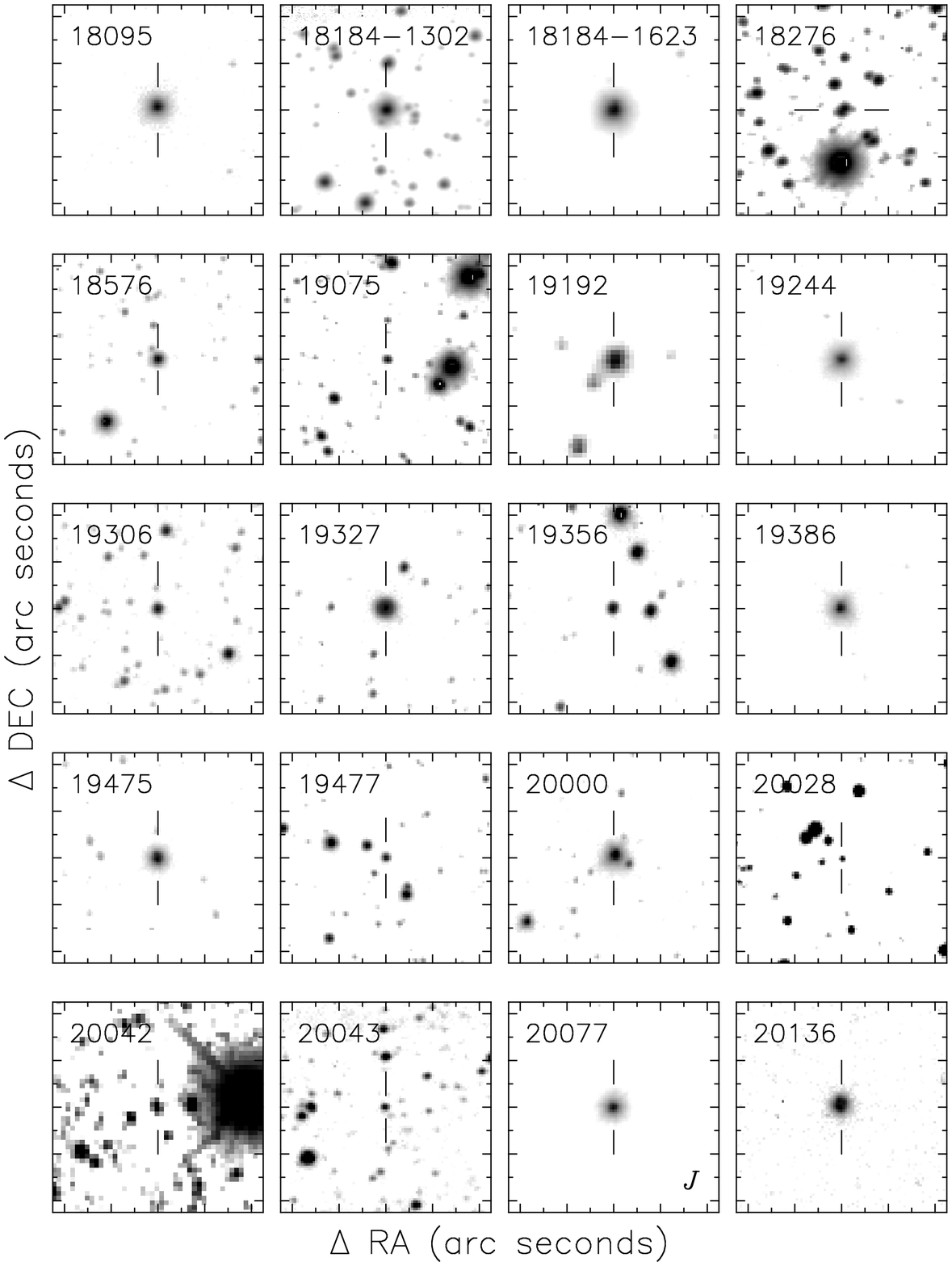}
\figcaption{%
Continued}
\end{figure}

\begin{figure}[p]
\plotone{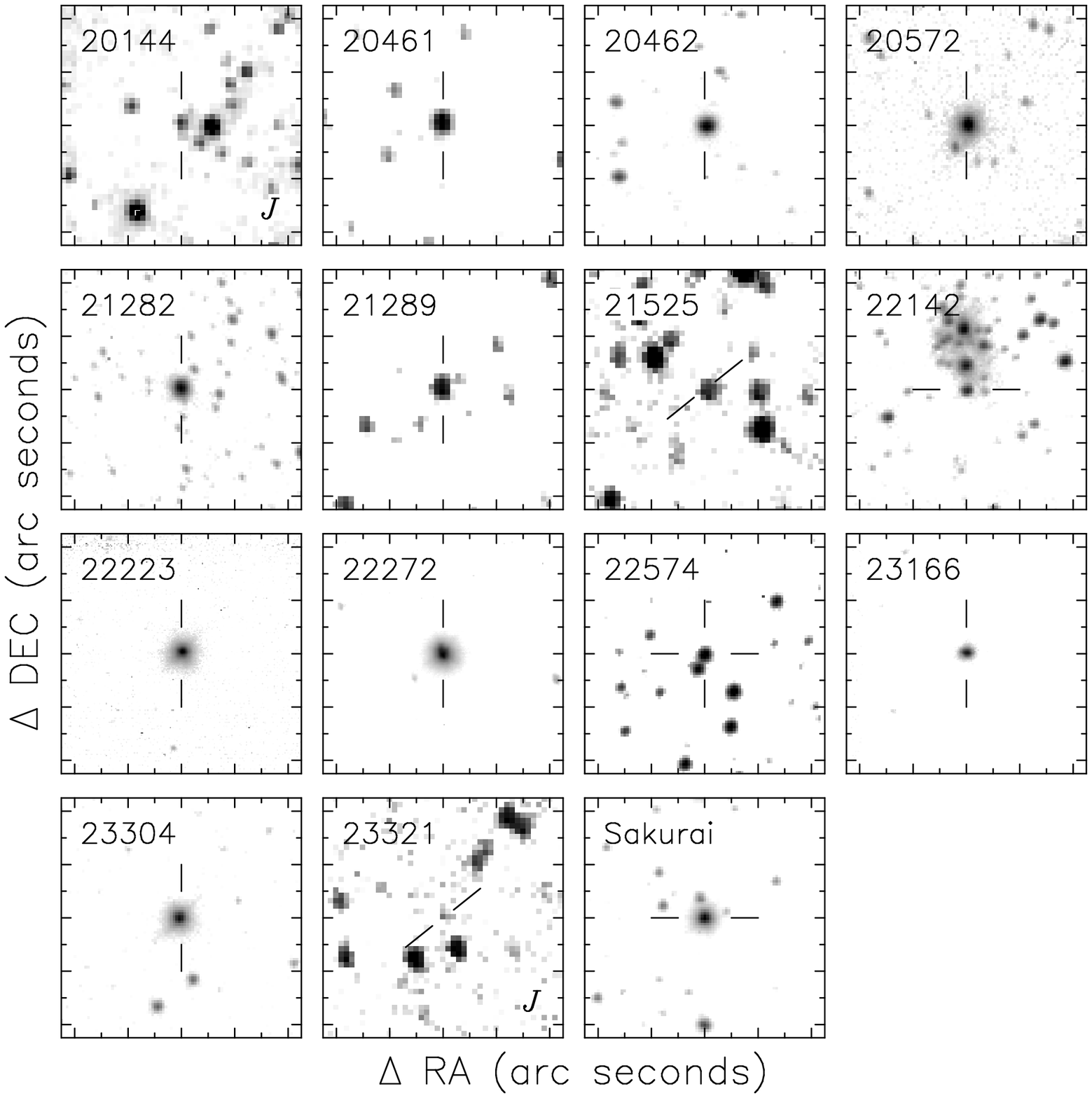}
\figcaption{%
Continued}
\end{figure}
\end{document}